\documentclass[trackchanges,twocolumn]{aastex701}

\begin{document}

\title{Deep Synoptic Array Science: Searching for Long Duration Radio Transients with the DSA-110}

\correspondingauthor{Myles B. Sherman}
\email{msherman@caltech.edu}

\author{Myles B. Sherman}
\affiliation{Cahill Center for Astronomy and Astrophysics, MC 249-17 California Institute of Technology, Pasadena CA 91125, USA.}
\email{msherman@caltech.edu}

\author{Nikita Kosogorov}
\affiliation{Cahill Center for Astronomy and Astrophysics, MC 249-17 California Institute of Technology, Pasadena CA 91125, USA.}
\email{nkosogor@caltech.edu}

\author{Casey Law}
\affiliation{Cahill Center for Astronomy and Astrophysics, MC 249-17 California Institute of Technology, Pasadena CA 91125, USA.}
\affiliation{Owens Valley Radio Observatory, California Institute of Technology, Big Pine CA 93513, USA.}
\email{caseylaw@caltech.edu}

\author{Vikram Ravi}
\affiliation{Cahill Center for Astronomy and Astrophysics, MC 249-17 California Institute of Technology, Pasadena CA 91125, USA.}
\affiliation{Owens Valley Radio Observatory, California Institute of Technology, Big Pine CA 93513, USA.}
\email{vikram@astro.caltech.edu}

\author{Jakob T. Faber}
\affiliation{Cahill Center for Astronomy and Astrophysics, MC 249-17 California Institute of Technology, Pasadena CA 91125, USA.}
\email{jfaber@caltech.edu}

\author{Stella K. Ocker}
\affiliation{The Observatories of the Carnegie Institution of Washington, Pasadena, CA 91101, USA.}
\affiliation{Cahill Center for Astronomy and Astrophysics, MC 249-17 California Institute of Technology, Pasadena CA 91125, USA.}
\email{socker@caltech.edu}

\author{Liam Connor}
\affiliation{Center for Astrophysics, Harvard \& Smithsonian, Cambridge, MA 02138, USA}
\email{liam.connor@cfa.harvard.edu}

\author{Yuanhong Qu}
\affiliation{Nevada Center for Astrophysics, University of Nevada, Las Vegas, NV 89154}
\affiliation{Department of Physics and Astronomy, University of Nevada Las Vegas, Las Vegas, NV 89154, USA}
\email{yuanhong.qu@unlv.edu}

\author{Kaitlyn Shin}
\affiliation{Cahill Center for Astronomy and Astrophysics, MC 249-17 California Institute of Technology, Pasadena CA 91125, USA.}
\email{kaitshin@caltech.edu}

\author{Kritti Sharma}
\affiliation{Cahill Center for Astronomy and Astrophysics, MC 249-17 California Institute of Technology, Pasadena CA 91125, USA.}
\email{kritti@caltech.edu}

\author{Pranav Sanghavi}
\affiliation{Center for Astrophysics, Harvard \& Smithsonian, Cambridge, MA 02138, USA}
\email{pranav.sanghavi@cfa.harvard.edu}


\author{Gregg Hallinan}
\affiliation{Cahill Center for Astronomy and Astrophysics, MC 249-17 California Institute of Technology, Pasadena CA 91125, USA.}
\affiliation{Owens Valley Radio Observatory, California Institute of Technology, Big Pine CA 93513, USA.}
\email{gh@astro.caltech.edu}

\author{Mark Hodges}
\affiliation{Owens Valley Radio Observatory, California Institute of Technology, Big Pine CA 93513, USA.}
\email{mwh@caltech.edu}

\collaboration{200}{(The Deep Synoptic Array team)}

\begin{abstract}

We describe the design and commissioning tests for the DSA-110 Not-So-Fast Radio Burst (NSFRB) search pipeline, a 1.4\,GHz image-plane single-pulse search sensitive to 134\,ms$-$160.8\,s radio bursts. Extending the pulse width range of the Fast Radio Burst (FRB) search by 3 orders of magnitude, the NSFRB search is sensitive to the recently-discovered Galactic Long Period Radio Transients (LPRTs {or LPTs}). The NSFRB search operates in real-time, utilizing a custom GPU-accelerated search code, \texttt{cerberus}, implemented in Python with JAX. We summarize successful commissioning sensitivity tests with continuum sources and pulsar B0329+54, estimating the {90\% completeness $25\sigma$} flux (fluence) {limit} to be {$\sim1200\,$mJy ($\sim160$\,Jy\,ms)}. Future tests of recovery of longer timescale transients, e.g. CHIME\,J1634+44, are planned to supplement injection testing and B0329+54 observations. An offline DSA-110 NSFRB Galactic Plane Survey was conducted to search for LPRTs, covering $-3.5^\circ<b<5.7^\circ$ and $141^\circ<l<225^\circ$ ($\sim770\,$square\,degrees) in Galactic coordinates. We estimate an upper limit Poissonian burst rate {$\sim2\,$hr$^{-1}$ per square degree ($\sim17$\,hr$^{-1}$} per $3^\circ\times3^\circ$ survey grid cell) maximized across the inner $|b|<0.25^\circ$ of the surveyed region. By imposing the {$\sim1200\,$mJy} flux limit on two representative models (the magnetar plastic flow model and the White Dwarf-M Dwarf binary model), we reject with 95\% confidence the presence of White Dwarf-M Dwarf binary LPRTs {(beamed in a detectable direction)} with periods between {$\sim10-50$\,s} within $\sim95\%$ of the surveyed region. Combined with the prevalence of LPRTs in the Galactic Plane, our results motivate further consideration of both White Dwarf-M Dwarf binary models and isolated magnetar models. We will continue to explore novel LPRT search strategies during real-time operations, such as triggered periodicity searches and additional targeted surveys.

\end{abstract}




\section{Introduction}\label{sec:introduction}


Radio pulsar searching is entering a new exploratory phase; traditional single-dish telescope searches such as the Pulsar Arecibo L-band Feed Array (PALFA), Parkes Multi-beam Survey (PMBS), and Five-Hundred Metre Aperture Spherical Telescope (FAST) have utilized multi-beam receivers to discover roughly 2/3 of the $\sim4000$ known Galactic pulsars \citep{manchester2005australia,manchester2001parkes,han2021fast,li2025searching}. Over the past decade, the TRAnsients and PUlsars with MeerKAT \citep[TRAPUM;][]{carli2024trapum}, the Canadian Hydrogen Intensity Mapping Experiment Pulsar \citep[CHIME/PULSAR;][]{dong2023second} backend, the LOw-Frequency ARray (LOFAR) Tied-Array All-Sky Survey \citep[LOTAAS][]{van2023lofar}, and Very Large Array (VLA) \texttt{realfast} system \citep{law2018realfast}, among others, have demonstrated that radio interferometers can achieve competitive sensitivities to single-dish telescopes with higher localization precision and less-costly antennas. Beyond contributing $\sim500$ additional pulsars, these arrays have been essential for follow-up imaging to identify and associate Supernova Remnants (SNRs), Pulsar Wind Nebulae (PWNe), long-duration afterglows, and Persistent Radio Sources (PRS) of both pulsars and extragalactic radio transients \citep[e.g.][]{gaensler2005expanding,bruni2025discovery,cotton2024meerkat}. However, these surveys have typically targeted pulsars with rotation periods $\lesssim 10$\,seconds out of necessity. Fast Fourier-domain search strategies require long integration times to remain sensitive to periods beyond second-scales, and suffer from low-frequency red noise, making them impractical targets \citep[][]{ransom2001new,singh2022gmrt}. Periodic transients with longer rotation periods (seconds to minutes), such as radio magnetars, Rotating Radio Transients (RRATs), and White Dwarf-M Dwarf binaries (either non-tidally-locked intermediate polars like AR Scorpii or tidally-locked non-accreting polars) were instead identified through serendipitous {single-pulse} surveys or follow-up of their high-energy counterparts \citep[e.g.][]{hyman2005powerful,caleb2022radio,marsh2016radio}.


The recent discovery of a dozen Galactic radio sources with unusually long pulse periods between $\sim20$\,seconds$-8.7$\,hours and low $\lesssim10\%$ duty cycles have begun to reveal an abundant unconstrained parameter space \citep[e.g.][]{hurley2022radio,de2025sporadic,horvath2025unified,men2025highly,tan2018lofar}. These so-called ``Long-Period Radio Transients'' (LPRTs, or LPTs) exhibit some properties characteristic of neutron stars, such as high polarization fractions, polarization position angle (PA) swings, temporal sub-structure, and un-detectable optical/IR (OIR) emission \citep[][]{mitra2023meterwavelength,kramer2024quasi,1969ApL.....3..225R,chrimes2022new}. However, at least two LPRTs, ILT\,J1101+5521 ($P=2.1$\,hours) and GLEAM-X\,J0704-37 ($P=2.9$\,hours) reside in binaries with M-dwarf companions, with blue optical excess suggesting they are White Dwarfs \citep[][]{de2025sporadic,hurley20242}. While the remaining sample have no clear M-Dwarf companions, the recent discovery of a secondary $P=8.74$\,hr period of GPM\,J1839-10 solely from radio timing suggest{s} the White-Dwarf-M-Dwarf interpretation may extend to OIR-quiet candidates \citep{horvath2025unified,men2025highly}. Nonetheless, a plethora of both neutron star and White Dwarf-M Dwarf binary models have been proposed to explain these new phenomena \citep[e.g.][]{beniamini2023evidence,beniamini2020periodicity,cooper2024beyond,qu2025magnetic}.

LPRTs' rapid emergence has spurred the development of novel search strategies on an accelerated timescale. The GaLactic and Extragalactic All-sky Murchison Wide-field Array (MWA)-Extended (GLEAM-X) survey's first {discovery, GLEAM-X\,J1627-52, was found by comparing archival incoherently summed, low frequency ($72-231$\,MHz) visibility data from 2 months apart \citep{hurley2022radio,2025PASA...42..129H}}. These were followed by MeerKAT's (1284\,MHz) PSR\,J0901-4046, LOFAR's (135\,MHz) PSR\,J0250+5854, CHIME/FRB's (400-800\,MHz) CHIME\,J0630+25 and CHIME J1634+44, all of which were discovered as single-pulses in second and millisecond resolution beamformed visibilities \citep{caleb2022discovery,dong2025chime2,dong2025chime,tan2018lofar}. These beamformed discoveries resulted in higher resolution ($\sim$arcsecond-to-arcminute) initial localizations, enabling immediate radio and multi-wavelength follow-up. In addition, ASKAP's and MeerKAT's L-band observations confirmed that LPRTs are detectable and discoverable at higher frequencies, where propagation effects like scattering in the Interstellar Medium (ISM) are less pronounced {\citep[e.g.][]{cordes1986refractive, ocker2024ne2001p,jing2025fast}}. Notably, all LPRT discoveries have been via single-pulses or sub-pulses rather than folding on trial rotation periods\footnote{Follow-up observations have used periodicity search and timing techniques for confirmation, but no LPRTs have been \textit{discovered} initially using periodic search techniques.}.

Perhaps most promising is the discovery of four ASKAP LPRTs (ASKAP\,J1755-2527, ASKAP\,J1935+2148, ASKAP\,J1839-0756, ASKAP/DART-J1832-0911), LOFAR's ILT\,J1101+5521,  {and the MWA's GPM\,J1839-10} and GLEAM-X\,J0704-37 using second-duration radio image differencing pipelines \citep{wang2025detection,li202444,lee2025emission,caleb2024emission,de2025sporadic,mcsweeney2025new,dobie2024two,hurley2023long}. Radio imaging consists of binning the U-V (East-West and North-South coordinates of each antenna pair's separation) visibility coordinates on a grid and applying a 2-dimensional FFT. This is much faster than visibility beamforming and is mathematically identical, though some resolution and flux sensitivity are lost via the gridding process. The image-plane LPRT detections demonstrate that this sensitivity loss is negligible for LPRT timescales (seconds-to-minutes), motivating the development of real-time LPRT image plane search pipelines {like the Commensal Real-time ASKAP Fast-Transients Coherent (CRACO) Upgrade \citep{wang2025craft}, in addition to upcoming beamformer-based searches such as the CHIME/SLOW pipeline (J. Huang et al., in prep.).}


In this paper, we introduce the Deep Synoptic Array's (DSA-110) radio telescope's Not-So-Fast Radio Burst (NSFRB)  pipeline, an image-plane single-pulse search for LPRTs on 0.134-160.8\,second timescales. The NSFRB system builds on the existing real-time Fast Radio Burst (FRB) search, which has discovered $\sim60$ FRBs and localized $\sim30$ to host galaxies with arcsecond-scale precision \citep{ravi2023deep,law2024deep,sharma2024preferential}. In Section~\ref{sec:design} we describe the NSFRB system design and search pipeline; {in Section~\ref{sec:commissioning} we describe the commissioning NSFRB injection tests and observations of pulsar B0329+54;} in Section~\ref{sec:ngps} we report the results of a pilot DSA-110 NSFRB Galactic Plane Survey (DN-GPS); in Section~\ref{sec:discussion} we discuss the sensitivity limits placed on LPRT emission in the Galactic Plane, and its implications for the White Dwarf binary model; in Section~\ref{sec:conclusion} we conclude with next steps, specifically the real-time search which is currently underway.

\section{NSFRB System Design}\label{sec:design}

\begin{figure*}
    \centering
    \includegraphics[width=\textwidth]{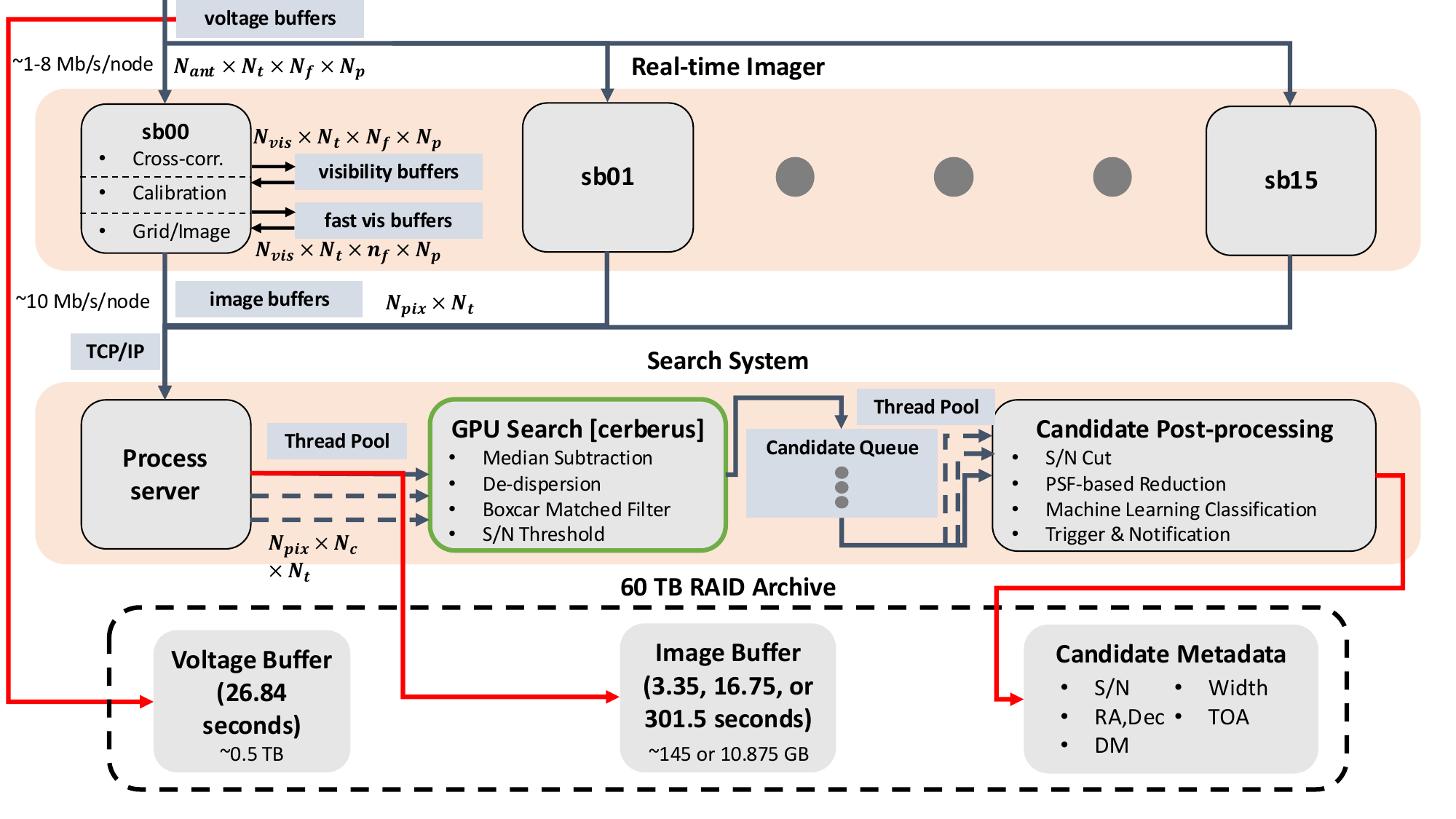}
    \caption{Diagram of the NSFRB Search System. Voltages for $N_{\rm ant}=97$ antennas, $N_t=25$ time samples, $N_p=2$ polarizations, and $N_f=384$ frequency channels are pulled from a rotating PSRDADA buffer, cross-correlated, down-channelized to $n_f=8$, and calibrated. $N_{\rm pix}=301\times301$ images are formed for each of $N_c=16$ sub-bands and sent to the Process Server which assigns search and post-processing tasks. Image data and candidate metadata are saved to disk along with 26.8-second of voltage data triggered from the PSRDADA buffer if candidates above the $6\sigma$ threshold are identified. We outline in green the post-processor which is \textit{not} required to operate in real-time.}
    \label{fig:nsfrb}
\end{figure*}

The DSA-110 is a 96-antenna drift-scan {\citep[see e.g.][for examples of drift-scan transient surveys with the MWA]{mantovanini2025galactic,2015PASA...32...25W}} radio interferometer at the Owens Valley Radio Observatory (OVRO) in Big Pine, CA. Details on the 64-antenna deployment and commissioning FRB survey can be found in \citet{ravi2023deep}. The array was completed in early 2025 by supplementing the East-West row of 47 ``core" antennas with a 35-antenna North-South arm, enabling real-time localization of radio sources without incorporating the 14 long-baseline ``outrigger" antennas. The antenna design, RF signal chain, baseband processing, and general compute resource layout remain largely similar to the 64-antenna deployment, and will not be detailed here. The revised visibility correlator, bandpass calibration procedures, and the FRB search pipeline will be discussed in detail elsewhere (Ravi et al., in prep.) For this work, we focus on components specific to the NSFRB system. Code for the NSFRB search is public at \url{https://github.com/dsa110/dsa110-nsfrb}, all of which is implemented using Python, leveraging the JAX\footnote{\url{https://docs.jax.dev/en/latest/index.html}} library to deploy GPU tasks. An ETCD\footnote{\url{https://etcd.io/}} server is used to manage alerts and coordination between compute nodes. Figure~\ref{fig:nsfrb} shows the full real-time system diagram, which we describe in the next sub-sections.

\subsection{Fringe-Stopping and Imaging}

The 187.5\,MHz observing band, centered on 1.4\,GHz, is split into $16\times11.7$\,MHz sub-bands, each processed on a single compute node. Each node's correlator outputs dual-polarization $N_{\rm base}=4656$ cross-correlated visibilities as $384\times30.5$\,kHz Nyquist-sampled ($32.7\mu$s) channels to a PSRDada buffer \citep{van2021psrdada}. For the NSFRB search, these are calibrated and fringe-stopped to the current meridian position in $\sim 3.35$\,second (102400 samples) chunks, down sampling to $134\,$ms resolution (25 samples) and $8\times1.46$\,MHz channels. The 3.35\,s fringe-stopping interval is selected to maintain point-source coherence; the fringe-timescale of the array is:

\begin{equation}
   t_F = 648000 {\rm s} \frac{ \lambda_c}{\pi b_{\rm max}15\cos{(\delta)}} \approx \frac{1.05\,{\rm s}}{\cos{(\delta)}}
\end{equation}

\noindent for the DSA-110, $b_{\rm max}=2.6\,$km, $\lambda_c=20$\,cm, where $\delta$ is the declination \citep[e.g.][]{thompson2017interferometry}. The array operates in drift-scan mode (no azimuthal slew) at $\delta=71.6^\circ$ during normal operations, making the nominal fringe rate $t_{F0}\approx3.36\,$s. 

Fringe-stopped visibilities are written to a 14.9\,MB buffer read by the correlator node imager, which grids visibilities with baseline lengths $b>b_{\rm min}=20\,$m from the dense 82-antenna `core' ($b<b_{\rm max}=485\,$m) to a $175\times175$-pixel UV-grid. The core is assumed to be coplanar, so the W-term is neglected for normal operation; however, W-stacking is implemented and may be explored in the future \citep[e.g.][]{gheller2023high}. Outrigger antenna data is included only for detailed manual inspection of candidates. The two visibility polarizations are summed, then approximately uniform weighting \citep[Briggs robustness parameter $\mathcal{R}=-2$;][]{briggs1995high} is applied to accumulate visibilities, and a 2-dimensional inverse-FFT\footnote{The NumPy Python Package implementation of \texttt{np.fft.ifft2d} is used for imaging. The visibility grid is first shifted using \texttt{np.fft.ifftshift} to properly order the zero-frequency component; \texttt{np.fft.ifftshift} is then applied to the resulting image.} is applied to form the final image with $\Delta \alpha=10.3''(\cos{\delta})^{-1}\,,\,\Delta \delta= 10.3''$ pixel scale (3 pixels per $31''$ synthesized beamwidth). {The DSA-110 primary beam FWHM is roughly $\Omega = 1.5^\circ(\cos{\delta})^{-1}\times1.5^\circ$, while the $175\times175$ pixel image covers a $\Omega = 0.5^\circ(\cos{\delta})^{-1}\times0.5^\circ$ search region}. Each of the 8 channels are imaged individually, then summed together to form a single image for each $134\,$ms time sample. Images on each node are then buffered (6.125\,MB) for transport by an http client to a central search server.

\subsection{\texttt{cerberus} Search Pipeline}

Images are transferred from each correlator node to the search process server using TCP/IP via a 40\,Gb ethernet switch, where they are associated and ordered based on their Modified Julian Date (MJD). The real-time search imposes a 3.35\,second timeout to accommodate correlator node failure or high traffic from adjacent DSA-110 pipelines (e.g. the FRB search). The server manages and deploys search tasks on one of two NVIDIA GeForce RTX 2080 Ti GPUs, which are coordinated (concatenation, noise statistics, candidate identification, and post-processor queuing) with one of 15 reserved threads on the Intel(R) Xeon(R) Silver 4210 CPU.

The custom \texttt{cerberus} single-pulse search pipeline implements Python JAX Just-In-Time (JIT) compiled functions to de-disperse and boxcar filter images on the GPU. Data is first median subtracted in each pixel, then de-dispersed over 16 DM trials between $171-4000$\,pc\,cm$^{-3}$ (DM tolerance of 1.6). The JAX de-dispersion method is adapted from a custom PyTorch implementation, \texttt{PyTorchDedispersion}\footnote{\url{https://github.com/nkosogor/PyTorchDedispersion}} \citep[][]{2019arXiv191201703P,2025ApJ...985..265K}, and allows shift indices for each frequency channel to be pre-computed and loaded onto the GPU to minimize latencies. The resulting de-dispersed time series are boxcar filtered on $\log_2$ spaced trials from 1-16 samples ($0.134-2.14$\,s), then normalized by the running standard deviation to compute the S/N. Final candidates are found by taking the maximum S/N (and its TOA) over time.  A nominal $6\sigma$ signal-to-noise (S/N) threshold is imposed to identify candidates, which are pushed to the post-processor.

\begin{figure*}
    \centering
    \includegraphics[width=\textwidth]{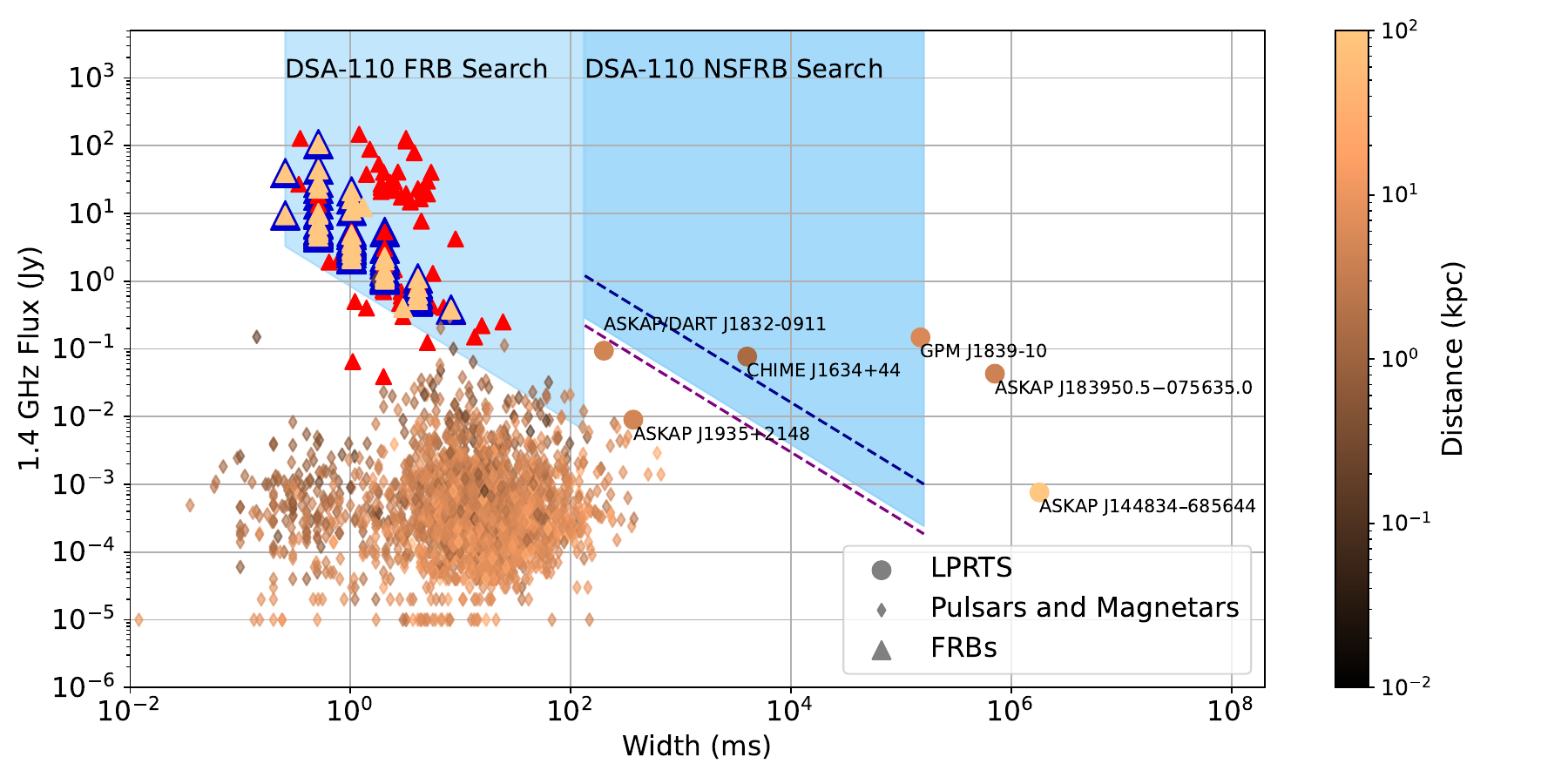}
    \caption{Pulse width and $6\sigma_{\rm srch}$ Fluence Sensitivity to $\Delta t =134$\,ms Bursts of the DSA-110 FRB (light blue; {$\sim 840\,$mJy\,ms}) and NSFRB (dark blue hatched; $\sim 40\,$Jy\,ms) searches. For each search we assume the fluence threshold is fixed; thus the lower border indicates the corresponding flux density threshold $6\sigma_{\rm th}\Delta t/W$ where $W$ is the width {measured at 1.4\,GHz (note that in some cases \citep[e.g. ASKAP\,J1832-09][]{li202444,wang2025detection}, only sub-pulses were detected at 1.4\,GHz while a broader profile was detected at lower frequencies. In these cases, we still use the 1.4 GHz sub-pulse width)}.  {The dark blue dashed line shows the 90\% completeness limit $25\sigma\approx1200$\,mJy ($25\sigma\Delta t\approx 160$\,Jy\,ms; see Section~\ref{sec:detrate}).} The purple dashed line shows the theoretical radiometer $6\sigma_{\rm th}$ sensitivity of the NSFRB survey. 1.4\,GHz fluences and widths are shown for radio pulsars in the ATNF catalog (diamonds), FRBs from the literature (triangles), and LPRTs (circles). DSA-110 FRBs have dark blue outlines and show their detection pulse widths and S/N rather than refined estimates from baseband analysis. The colorbar indicates distance estimates in kpc (FRB luminosity distances are computed from host Galaxy redshifts; pulsar, magnetar, and LPRT distances are from parallaxes where available, and DM distance limits for the rest of the sample). {Note that only pulsars, FRBs, and LPRTs with 1.4 GHz detections are shown.}}
    \label{fig:sens}
\end{figure*}

The search is performed at three sampling timescales: 0.134\,s, 0.67\,s (binned by 5 samples), and 10.05\,s (binned by 75 samples). The process server bins images at each timescale as they are received, aligning them in the RA direction to account for the 25-sample fringe-stopping interval, and preserves batches for \texttt{cerberus} to re-search at the 0.67\,s and 10.05\,s timescales. For the real-time search, a single search iteration on the next shortest timescale (0.134\,s, 0.67\,s) is dropped to enable the longer timescale  (0.67\,s, 10.05\,s) search iterations (the Galactic Plane data is not searched in real-time and thus does not drop iterations). The 0.67\,s data is searched identically to the 0.134\,s data, now probing DMs from $855-20000\,$pc\,cm$^{-3}$\,\footnote{While it is not strictly necessary to search up to $20000\,$pc\,cm$^{-3}$ since the peak DMs in the Galactic Plane typically only reach $\sim5000$pc\,cm$^{-3}$, this range was selected because (1) the coarse time and frequency resolution, there is little benefit from finely sampled DM trials, and (2) extending the DM range by exactly $5\times$ for data sampled at $5\times$ lower resolution allows the GPU search code to utilize the same pre-computed shifts, thus saving GPU compute time and memory by only loading the the lookup table once.} with widths from $0.67-10.72\,$s. The longest timescale data bypasses de-dispersion, leveraging the fact that the DM delay for DMs below $\sim20000\,$pc\,cm$^{-3}$ is now contained to a single sample across the 187.5\,MHz band. They are then searched from $10.05-160.8$\,s. The sensitivity of the search is shown in Figure~\ref{fig:sens}, which indicates the full range of single-pulse widths the NSFRB search probes. {The broad morphology of known LPRTs includes those with smooth profiles \citep[e.g.][]{mcsweeney2025new} and `spiky' sub-pulses \citep[e.g.][]{li202444,dong2025chime}. By searching at multiple timescales, \texttt{cerberus} will be sensitive to both morphologies.} 

\begin{figure*}
    \centering
    \includegraphics[width=\textwidth]{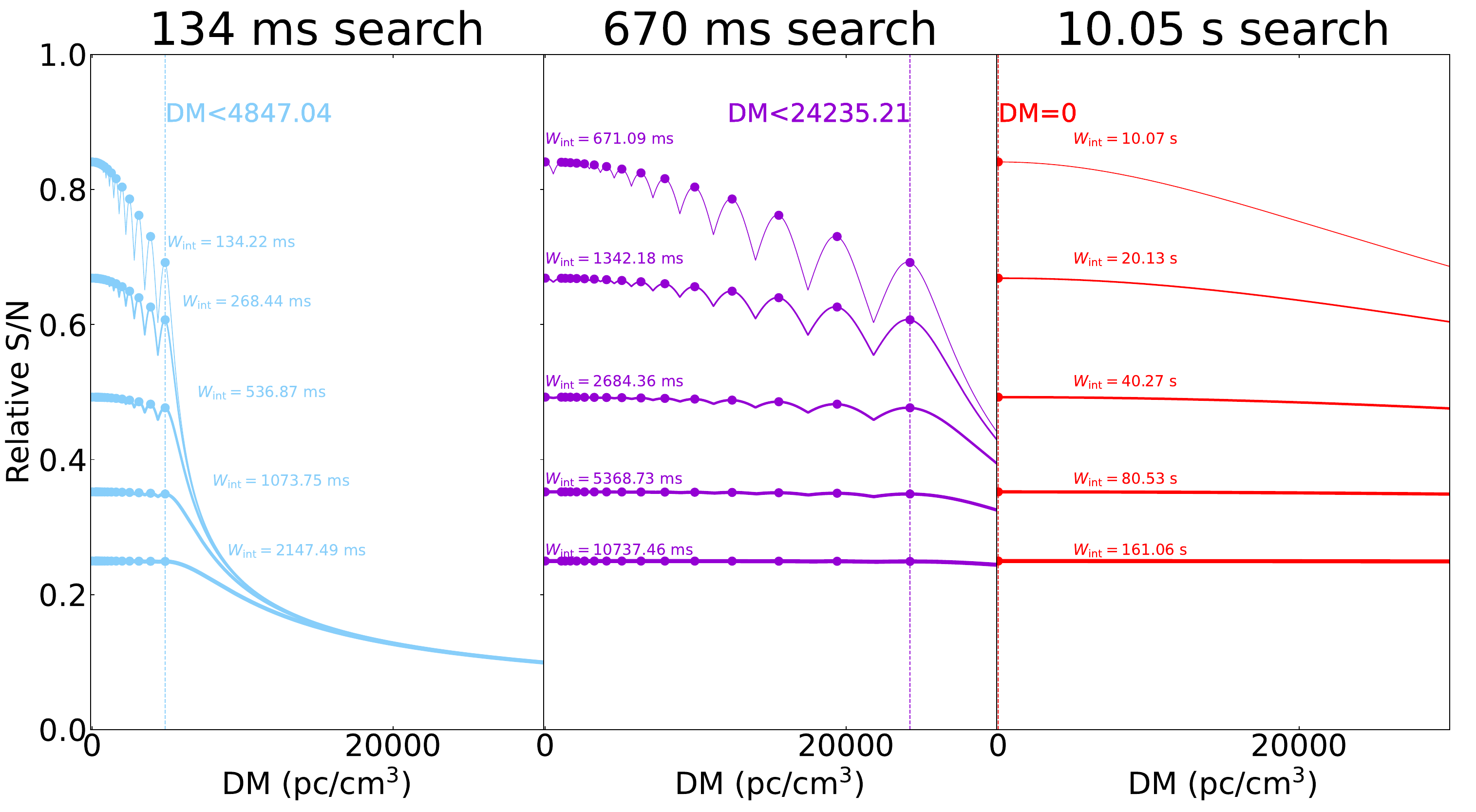}
    \caption{{Relative S/N Loss (Equation~\ref{eq:snr}) of NSFRB Search on the Three \texttt{cerberus} Timescales (134\,ms, 670\,ms, and 10.05\,s) as a Function of DM. Each line corresponds to an intrinsic pulse width $W_{\rm int}$, and solid points indicate each trial DM. }}
    \label{fig:dmsens}
\end{figure*}

{Figure~\ref{fig:dmsens} shows the sensitivity as a function of DM for the de-dispersion schemes on each search timescale. We compute the total smeared pulse width by summing in quadrature the intrinsic pulse width $W_{\rm int}$, search timescale $t_s$, the dispersive smearing in the lowest frequency channel $t_{\Delta\nu}=(0.0083\,{\rm ms})\Delta\nu_{\rm MHZ} DM\nu_{\rm min,GHz}^{-3}$, and the dispersive delay across the full bandwidth after de-dispersion to the nearest trial DM $t_{\rm BW}=(4.15\,{\rm ms})|DM-DM
_{\rm trial}|(\nu_{\rm min,GHz}^{-2}-\nu_{\rm max,GHz}^{-2}) $  . The S/N relative to DM=0 is then:}

{\begin{equation}\label{eq:snr}
    {\rm S/N_{\rm rel}}(DM)=\sqrt{\frac{\sqrt{W_{\rm int}^2 + t_s^2 + t_{\Delta  \nu}^2+t_{\rm BW}^2}}{t_s}}
\end{equation}}

\noindent {Thus ${\rm S/N_{\rm rel}}(DM)$ is the S/N relative to that of a pulse with negligible $W_{\rm int}$ at $DM=0$. Figure~\ref{fig:dmsens} shows ${\rm S/N_{\rm rel}}$ for each \texttt{cerberus} search timescale. We observe first that the relative S/N drops rapidly below $60\%$ on the 134\,ms timescale for $DM\gtrsim4000\,$pc\,cm$^{-3}$, and that pulses beyond $\sim540\,$ms (covered by the two longest trial widths) have $\lesssim40\%$ even at $DM=0$. This motivates the 670\,ms search, which preserves $\gtrsim60\%$ relative S/N for $670-1350$\,ms timescales while minimizing S/N loss at high DM trials. Finally, the S/N loss flattens between trials for long trial widths, allowing the 10.05\,s search to forego de-dispersion with minimal sensitivity loss while recovering the longest timescale bursts. Therefore, while binning to longer timescales loses raw fluence sensitivity as shown in Figure~\ref{fig:sens}, it preserves sensitivity to the longest pulses at high DM trials relative to $DM=0$.}

\subsection{Candidate Post-Processing}\label{sec:postproc}

When candidates are identified, the image, S/Ns for all pixels, widths, and DM trials, and times-of-arrival (TOAs) for all pixels, widths, and DM trials are saved to disk. The candidate timestamp is pushed to ETCD which is monitored by the post-processing system. Standard post-processing is limited to the 10 highest S/N candidates, under the assumption that each image will have at most one candidate. To identify the most likely candidate, we compare the candidates' relative positions within the image to a simulated PSF. First, the PSF is thresholded at the 90$^{\rm th}$ percentile to make a binary mask, which is centered on each candidate in turn. This candidate is treated as the central PSF lobe, and any other candidates within the binary mask are identified as possible sidelobes. The weighted mean S/N is computed over the central and sidelobe candidate S/Ns. After repeating with each candidate, the maximum S/N is identified and its central lobe candidate's width, DM, position, and TOA are taken as the final, most-likely candidate.

The second major post-processing stage is to apply a three-stage Convolutional Neural Network (CNN) classifier to the image {(centered on the TOA)} to identify and rule out Radio Frequency Interference (RFI). The CNN is trained on a set of simulated point sources, simulated near- and far-field RFI {(examples of RFI are shown in Appendix~\ref{app:rfi})}, and field data sets of RFI and pulsar B0329+54 from NSFRB commissioning. The model consists of successive convolutional blocks with batch normalization, nonlinear activations, and max-pooling, followed by a small fully connected head with dropout; {the CNN parameter learning rates are optimized with an Adaptive Moment Estimation \citep[\texttt{Adam};][]{adam2014method} optimizer using} a binary cross-entropy loss. After applying the trained model to the 3D image cube (position, time, frequency), the network outputs a single logit for each candidate which, after a sigmoid activation, we interpret as the RFI probability {$p_{\rm rfi}$} (0 indicates likely astrophysical; 1 indicates likely RFI). Simulated injection tests and pulsar observations were used to verify classifier performance. Positively identified candidates are assigned a candidate name and metadata (RA, DEC, TOA, DM, width, and S/N) are saved to a \texttt{json} file. A candidate plot is saved and pushed as a Slack alert for human inspection; this shows the time and frequency averaged image, the S/N as a function of trial DM and width for the final candidate pixel, the dynamic spectrum from the final candidate pixel, and the de-dispersed time series from the final candidate pixel. Finally, a trigger is sent to the correlator nodes to dump voltage data for a $26.85$\,second interval around the burst to disk for offline analysis and localization; voltage triggering is still being implemented. {Code for the NSFRB pipeline, including imaging, data transfer, searching, and post-processing, is publicly available at \url{https://github.com/dsa110/dsa110-nsfrb}.}


\section{Commissioning Injection and Pulsar Tests}\label{sec:commissioning}

NSFRB commissioning observations took place from February - October, 2025, including injection testing, pulsar observations, continuum source observations, and an offline Galactic Plane Survey (February - April). the latter two are discussed in Section~\ref{sec:ngps}.

\subsection{{Initial S/N Threshold Optimization}}\label{sec:initsnr}


{We conducted a simulated pulse injection test to estimate an optimal S/N threshold which could be tuned further throughout commissioning. First, }$\sim200$ boxcar-shaped pulses with DMs between $0-4000\,$pc\,cm$^{-3}$, widths between $0.134-2.14\,$s, and S/N between $0-10\sigma$ were generated. The amplitude of each burst is scaled based on noise estimates from imaged fast visibility data. These were sent to the process server and searched, with the peak S/N candidate identified and saved. {We then compute a false-negative rate (fraction of missed injections) as a function of trial S/N thresholds from 0 to 20.} $\sim2000$ noise realizations without pulses were similarly sent through the search pipeline to estimate the false-positive rate {as a function of S/N threshold. We simultaneously minimize the false-positive and false-negative rates with respect to S/N threshold, obtaining an optimal $6\sigma$ detection limit. This} $6\sigma$ threshold was used for the Galactic Plane survey, while a more stringent $8.5\sigma$ has been adopted for the real-time pipeline to limit false positives while commissioning observations are completed. {The threshold for the real-time search will be further tuned as we better characterize the RFI environment and noise statistics.}

\subsection{{Detection Rate Injection Tests}}\label{sec:detrate}

\begin{figure}
    \centering
    \includegraphics[width=\linewidth]{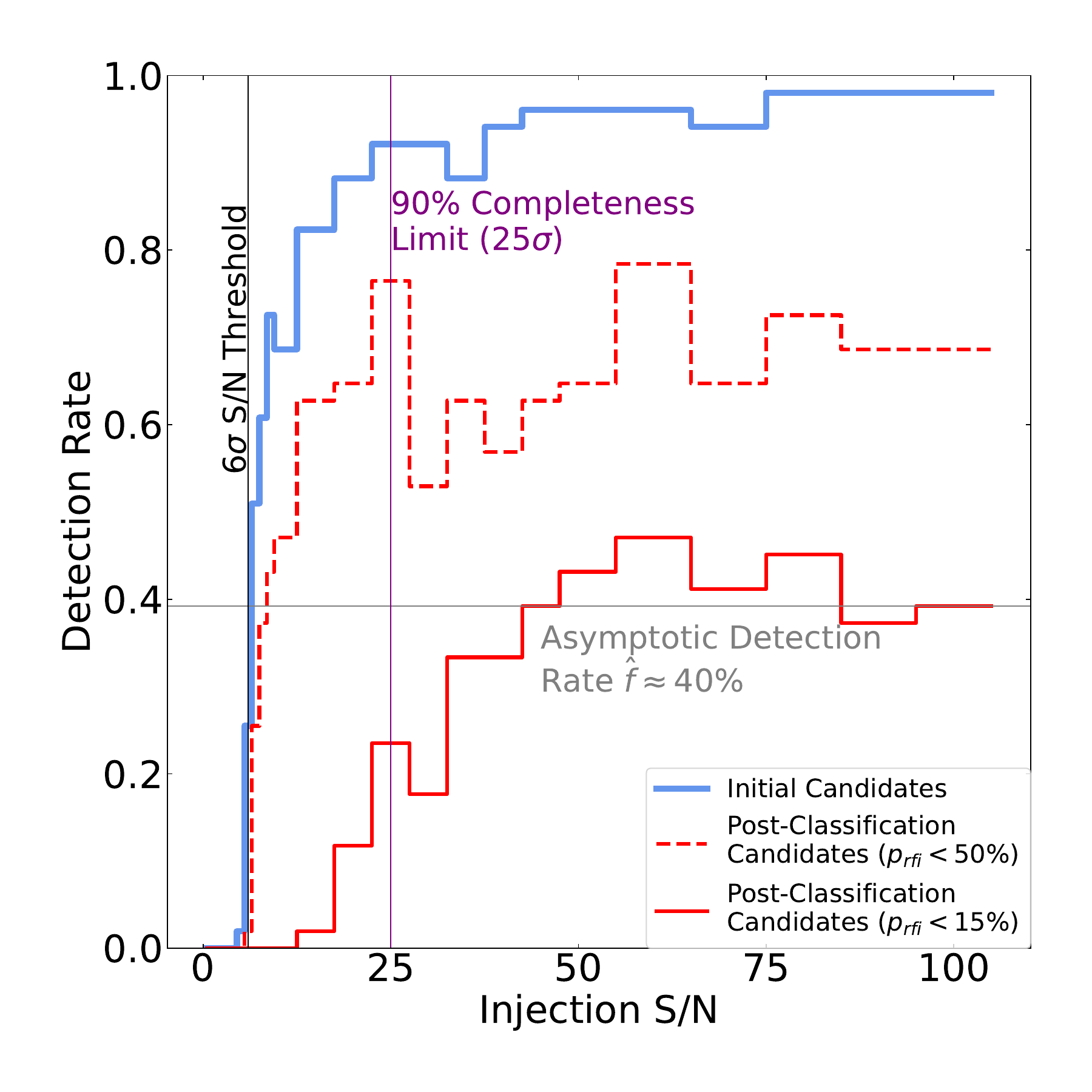}
    \caption{{Detection Rate vs. Input S/N for Injection Test. The blue curve includes any candidates detected with S/N$>6\sigma$ while the red curves include only candidates classified as sources by the CNN classifier using probability thresholds $p_{\rm rfi}<50\%$ (dashed) and $p_{\rm rfi}<15\%$ (solid). The black line indicates the $6\sigma$ threshold determined in Section~\ref{sec:initsnr}. }}
    \label{fig:finalinjections}
\end{figure}

{We repeat the simulated injection test using the $6\sigma$ threshold and the complete post-processing pipeline, generating 50 injections per S/N bin between $1-40\sigma$ at randomly selected trial DMs and pulse widths. Figure~\ref{fig:finalinjections} shows the resulting detection rate, computed as the fraction of recovered injections in each S/N bin. The blue curve includes any candidates detected with S/N$>6\sigma$; from this we estimate the search is $>90\%$ complete for S/N$>25
\sigma$. In Section~\ref{sec:results} we use this as the completeness flux limit for the NSFRB search. }

{As expected and shown by the red curves in Figure~\ref{fig:finalinjections}, the classifier reduces the search completeness by assessing each injection based on factors other than S/N (e.g. frequency extent, similarity to the point-spread function). The nominal search uses an RFI probability threshold $p_{\rm rfi }<15\%$ for the CNN to reduce false-positives; for this we estimate an asymptotic detection rate of $\hat{f}\approx40\%$. In Section~\ref{sec:results}, we incorporate this factor to estimate Poissonian burst rate limits.} 

{Raising the threshold to $p_{\rm rfi}<50\%$ recovers more injections but is less effective rejecting RFI. As new data is incorporated into the training set in future surveys, we expect this issue to improve gradually. The injection test also indicates some limitations of the search pipeline: 
\begin{itemize}
    \item The instantaneous noise estimate can be biased high for wider injections (8-16 samples) which lowers some candidates' S/N;
    \item Injections for which the burst (lowest frequency channel) is not at 50\% phase are often rejected, which is mitigated by the centering method (see Section~\ref{sec:postproc}) but will limit detection of bursts with the incorrect TOA;
    \item In a small fraction of cases, the PSF clustering technique identifies ``off-axis" candidates as opposed to the central candidate; further characterization of the clustering pipeline is needed to understand what properties cause this.
\end{itemize} 
\noindent These limitations will be investigated in future work.}

\subsection{Pulsar Observations}

\begin{figure*}
    \centering
    \includegraphics[width=\textwidth]{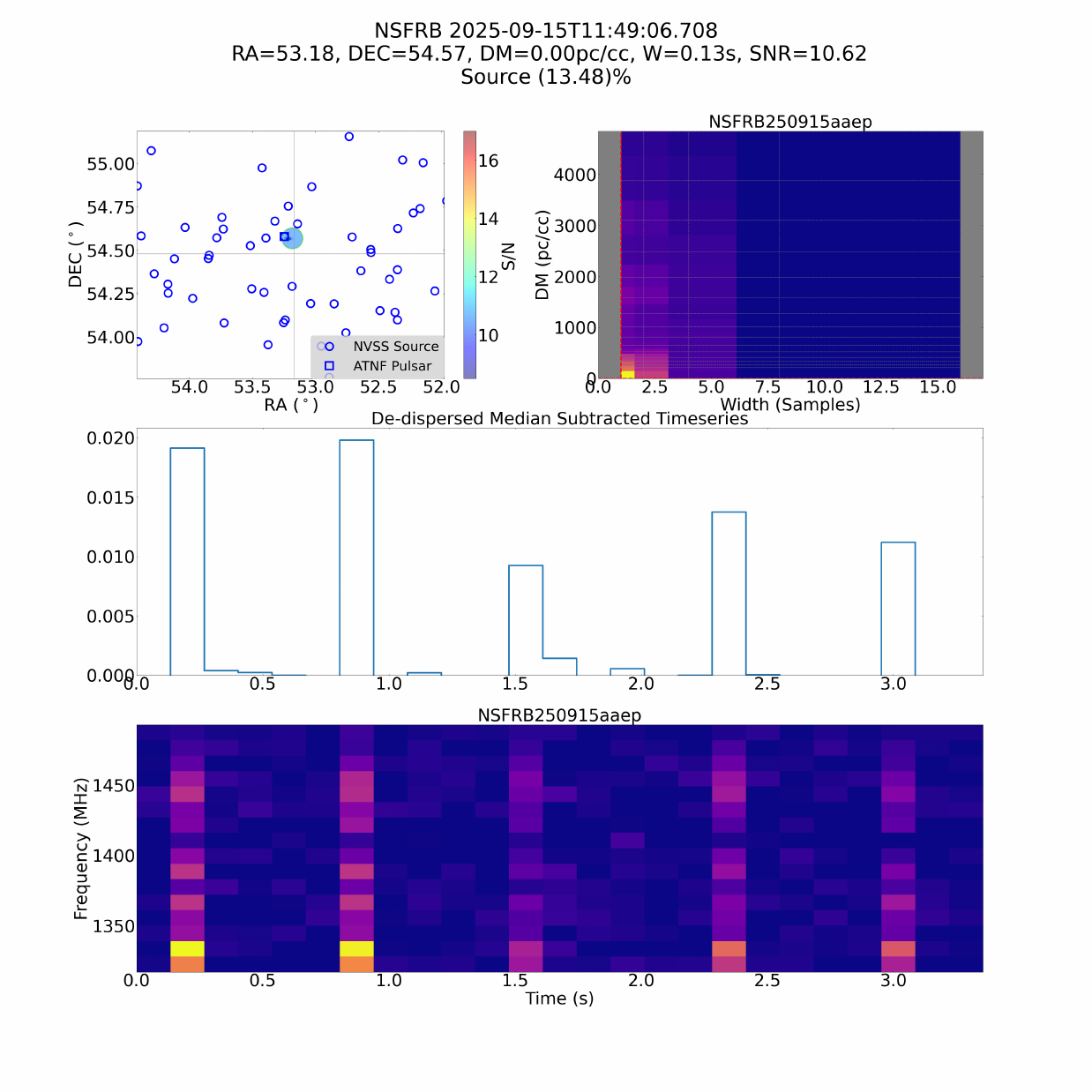}
    \caption{Example NSFRB candidate plot from detections of pulsar B0329+54. (top left) Position of the candidate (shaded circle) in reference to the pulsar's catalogued position (open blue square) and other field NVSS continuum sources (open blue circles). The colorbar corresponds to the S/N. (top right) S/N as a function of search DM and pulse width. (middle) DM=0 0.134\,ms-sampled time series of the peak candidate in arbitrary units. (bottom) Dynamic spectrum of the peak candidate with coarse 12\,MHz channels. The title gives the UTC time of the first time sample, RA and declination after systematic error correction, detected DM, pulse width, and S/N, and the CNN classifier's p-value.}
    \label{fig:b0329}
\end{figure*}

Observations of pulsar B0329+54 \citep[$P=0.7145$\,s, pulse profile FWHM$=6.6\,$ms, $S_{\rm 1.4\,GHz}=203$\,mJy][]{manchester2005australia} were used to monitor the system's sensitivity to transient pulsed emission which the NSFRB search is targeting. This pulsar is selected for its high pulse period ($\sim5\times134$\,ms samples), long pulse duration, and high flux; it is the only northern sky pulsar with an expected single-pulse S/N$>5\sigma_{\rm srch}$. Single pulses (filling the remaining samples with Gaussian noise) from initial observations were used to train the CNN classifier. Later observations then properly detected and classified the pulsar. Figure~\ref{fig:b0329} shows a candidate summary plot in which {one of the five visible B0329+54 pulses was detected}, demonstrating that the NSFRB search is sensitive to both single-pulsed and periodic emission. Note that pulsar\,B0329+54 has DM$=26.76410\,$pc\,cm$^{-3}$, but was detected at the DM=0\,pc\,cm$^{-3}$ trial bin due to the coarse trial DM spacing. The next DM trial is $213.72\,$pc\,cm$^{-3}$.

\begin{figure*}
    \centering
    \includegraphics[width=\textwidth]{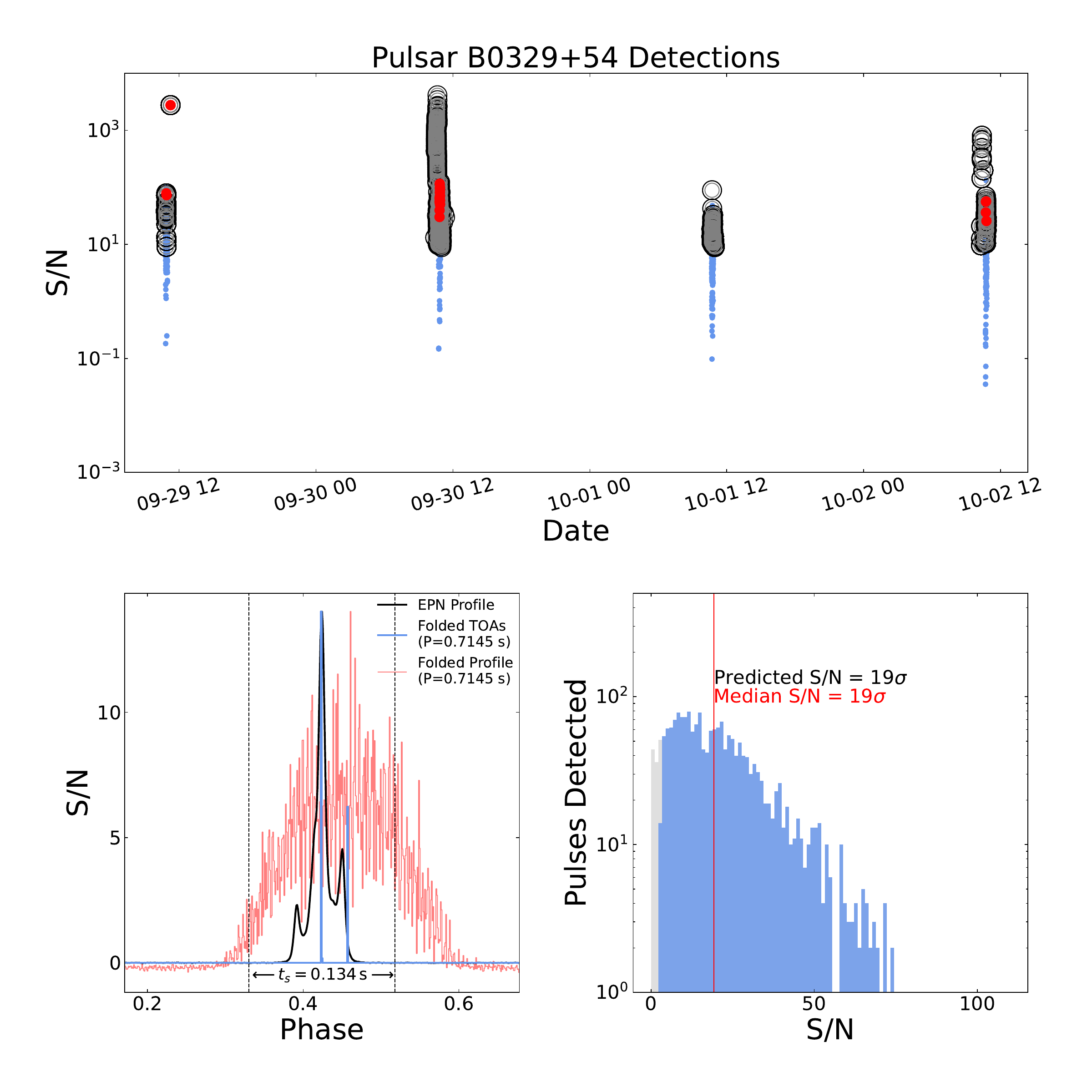}
    \caption{Summary of Pulsar B0329+54 Monitoring Observations. (Top) The S/N of all pulses measured from predicted Times-of-Arrival (TOAs) are in blue. Initial candidates above the $8.5\sigma$ threshold are in grey, remaining candidates after PSF-based reduction are in black, and final candidates after classification are in red. (Bottom right) The S/N distribution of all single-pulses above $3\sigma$ are in blue, while `non-detections' below $3\sigma$ are in grey. The median of the $>3\sigma$ distribution is indicated in red and compared to the predicted S/N in black. (Bottom left) {The folded pulse profile (S/N-weighted histogram of full timeseries) with 1036 phase bins is shown in red while a histogram of peak TOAs is shown in blue.} Note that the folded profile {TOAs are} not phase connected between days, nor {are they} phase connected to past observations \citep[e.g.][]{deller2019microarcsecond}. The profile from the European Pulsar Network (EPN) database is shown in black \citep[][\url{https://psrweb.jb.man.ac.uk/epndb}]{seiradakis1995pulsar}. {The TOA histogram is normalized such that the sum equals the mean TOA S/N, while the EPN and folded profiles are scaled to match the TOA histogram peak. The sampling time 134\,ms, which limits the pulse profile resolution, is shown in phase units as black dashed lines. }}
    \label{fig:b0329all}
\end{figure*}

The pulsar was observed during its transit for four consecutive days (MJDs 60947-60950; September 29 - October 2, 2025) to characterize the search sensitivity at each stage of the pipeline. Figure~\ref{fig:b0329all} summarizes the results; {2039} total pulses were recorded, with {1863} detected above $3\sigma_{\rm srch}$. {The search is limited to report only one candidate per 3.35\,s timeseries; therefore, we expect $2039\times (134\,{\rm ms}/0.7145\,{\rm s})\approx382$ final candidates from the 2039 total pulses.} 8621 total candidates above the $8.5\sigma$ real-time threshold were detected; PSF-based reduction resulted in 1511 candidates; the classifier then allowed through {33} final candidates for which alerts were sent{, a roughly 9\% detection rate}. No final candidates were identified on 2025-10-01; inspection of the data implies this was due to higher-than-average RFI contamination and an unfortunately timed system restart. {From the TOA histogram, it appears that the two brightest peaks of the B0329+54 profile are detected while the third is not, though we do not perform rigorous timing; the full profile shape is in part limited by the sampling time.} The S/N distribution of pulses $>3\sigma$ peaks near {$19\sigma$ which is consistent expectations from the known 1.4\,GHz flux and NSFRB sensitivity}. The B0329+54 observations demonstrate the pipeline has sensitivity aligned with theoretical predictions, can decimate a large number of triggers to a manageable number of candidates, and can properly classify real astrophysical single-pulses.


\section{The DSA-110 NSFRB Galactic Plane Survey (DN-GPS)}\label{sec:ngps}

\subsection{Survey Strategy}

\begin{figure*}
    \centering
    \includegraphics[width=\textwidth]{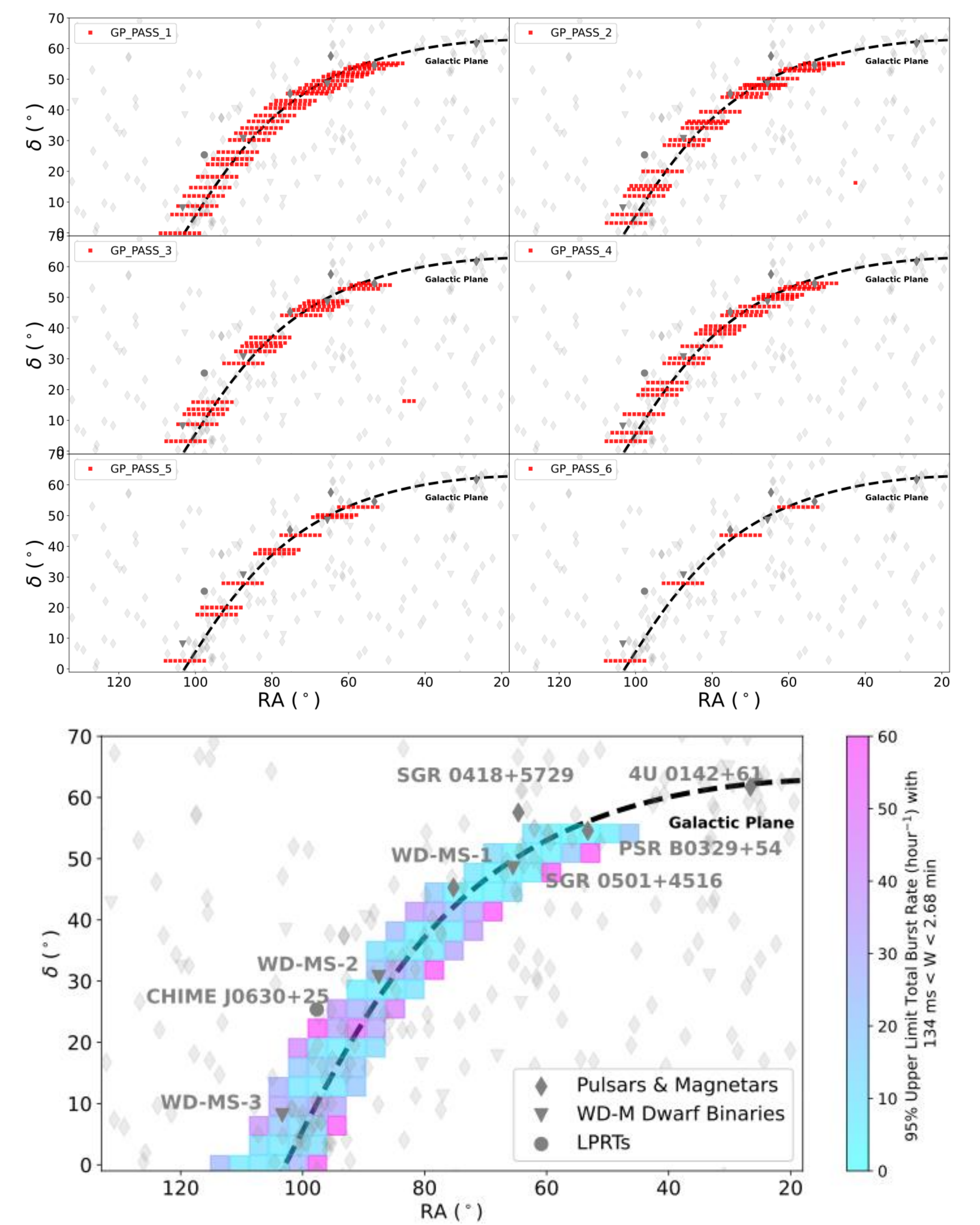}
    \caption{Coverage and Rate Limits for the DN-GPS pilot survey. (top) Each sub-plot shows one of 6 passes of the Galactic Plane (dashed black line). Center coordinates for each 3.35\,s pointing are shown in red. The positions of known pulsars and magnetars are shown as diamonds; non-pulsar LPRTs detected to date are circles. A sample of White Dwarf-M Dwarf binaries from \citet{2021MNRAS.506.5201R} are shown as triangles. (bottom) 95\% Poissonian burst rate upper limits are indicated by the colorbar for each $3^\circ\times3^\circ$ grid region as discussed in Section~\ref{sec:ngps}. Sources of interest covered by the DN-GPS search region are labeled; of these, only B0329+54 is confidently detected. }
    \label{fig:burstrate}
\end{figure*}

The NSFRB instrument conducted an inaugural survey of the Galactic Plane (GP) from February 18 - April 23, 2025 (MJD 60724 - 60788). {The DSA-110 is a `drift-scan' telescope, meaning that it cannot slew in the azimuthal direction. Visibilities are instead fringe-stopped to the meridian in 3.35\,second ($25\times134\,$ms sampled) integrations as the sky drifts overhead. To track the Galactic Plane, antennas were slewed in elevation every hour, initially pointing south of the plane and recording data as the plane transited across the field-of-view. This method enables roughly 10 degrees coverage above and below the plane on each slew.} This was repeated approximately daily starting 3 degrees further south each day, but slewing by the same hourly step-size. This strategy maximizes the time between slews (roughly hourly), extends vertical plane coverage, and limits instrumental strain by spacing its observation over a 12-day span. The 12-day pattern was repeated 6 discrete times to ensure complete coverage of the plane in the presence of maintenance or observer error. Figure~\ref{fig:burstrate} shows the DN-GPS coverage maps for each pass; the final survey covers $-3.5^\circ<b<5.7^\circ$ and $141^\circ<l<225^\circ$ in Galactic coordinates.

\subsection{Astrometric Calibration}

Astrometric errors for candidates from the DN-GPS are estimated by comparing measured positions of sources from the Radio Fundamental Catalogue (RFC) to their milliarcsecond scale Very Long Baseline Interferometric (VLBI) localizations \citep[][]{petrov2025radio}. 16 RFC calibrators are selected as those that have 1.4\,GHz flux measurements from the National Radio Astronomy Observatory (NRAO) VLA Sky Survey\footnote{The NVSS survey is used because the VLA D-configuration synthesized beamsize ($\sim46''$) is comparable to the DSA-110 core's ($\sim 31''$). Therefore, unresolved NVSS sources are most likely unresolved to the DSA-110, making flux measurements within a synthesized beam comparable.} \citep[NVSS;][]{condon1998nrao} and fall within the DN-GPS field (see Figure~\ref{fig:burstrate}). We form $175\times175$ pixel images around each calibrator transit (between $1-30$ images, each averaged over 3.35\,s) and use the \texttt{casatools} and \texttt{Astropy.WCS} modules to compute the RA and declination of each image pixel based on the current telescope pointing. Each image is fit with a 2D Gaussian around the main lobe of the RFC source to measure its position and align the images with each other, averaging over the full image set to maximize the signal-to-noise. This `alignment solution' - the number of image pixels by which the source transits between each 3.35\,s-fringestopped image - is also used to align images in the 0.67\,s and 10.06\,s search pipelines to correct small deviations from the expected transit due to the time and UV-grid resolution.

The final source position is measured from a 2D Gaussian fit of the full averaged image. We identify a statistically significant systematic position offset of $\epsilon_{\rm \alpha}=-1.77'\pm 2''$, $\epsilon_{\rm \delta}=-1.15'\pm1''$ in RA and declination, respectively. After applying this correction, the remaining RMS position error is $\sigma_{\alpha}=12''$, $\sigma_\delta=7''$. Figure~\ref{fig:calplot} (right) shows the final position offsets of the 16 calibrators with the $3\sigma$ position error circle.

\subsection{Sensitivity and Flux Calibration}

\begin{figure*}
    \centering
    \includegraphics[width=\linewidth]{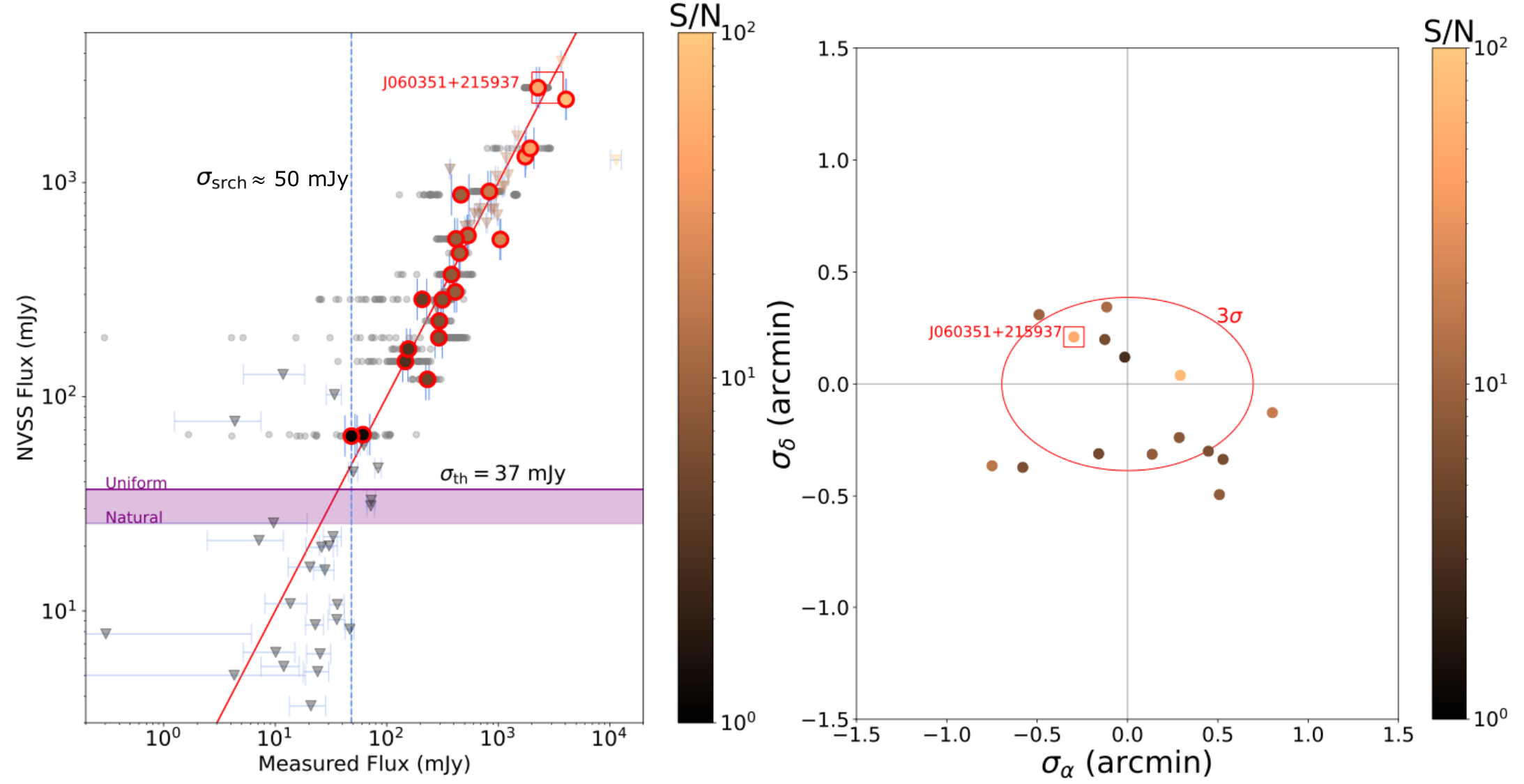}
    \caption{Flux and Astrometry Calibration Observations from the DSA-110 NSFRB Galactic Plane Survey (DN-GPS). (Left) Measured vs. NVSS catalogued flux for the 16 calibrators selected from DN-GPS data are shown as circles. Individual measurements are in grey, while the mean measurements have colorbar equal to the observed signal-to-noise. Additional calibrator observations from the real-time pipeline are shown as triangles. The red line shows the conversion curve constrained from the brightest source, J060351+215937, and the sensitivity limit $\sigma_{\rm srch}$ is shown in blue. The range of theoretical sensitivity limits from natural to approximately uniform $\mathcal{R}=-2$ imaging weights is shown in purple. ({Right}) Measured position errors of the 16 calibrators, after subtracting the mean systematic offsets $\epsilon_{\rm \alpha}=-1.77'\pm 2''$, $\epsilon_{\rm \delta}=-1.15'\pm1''$ from RFC positions, are shown as circles with colorbar equal to the observed signal-to-noise. The $3\sigma$ error ellipse ($72''\times42''$) is outlined in red.}
    \label{fig:calplot}
\end{figure*}

\begin{figure}
    \centering
    \includegraphics[width=\linewidth]{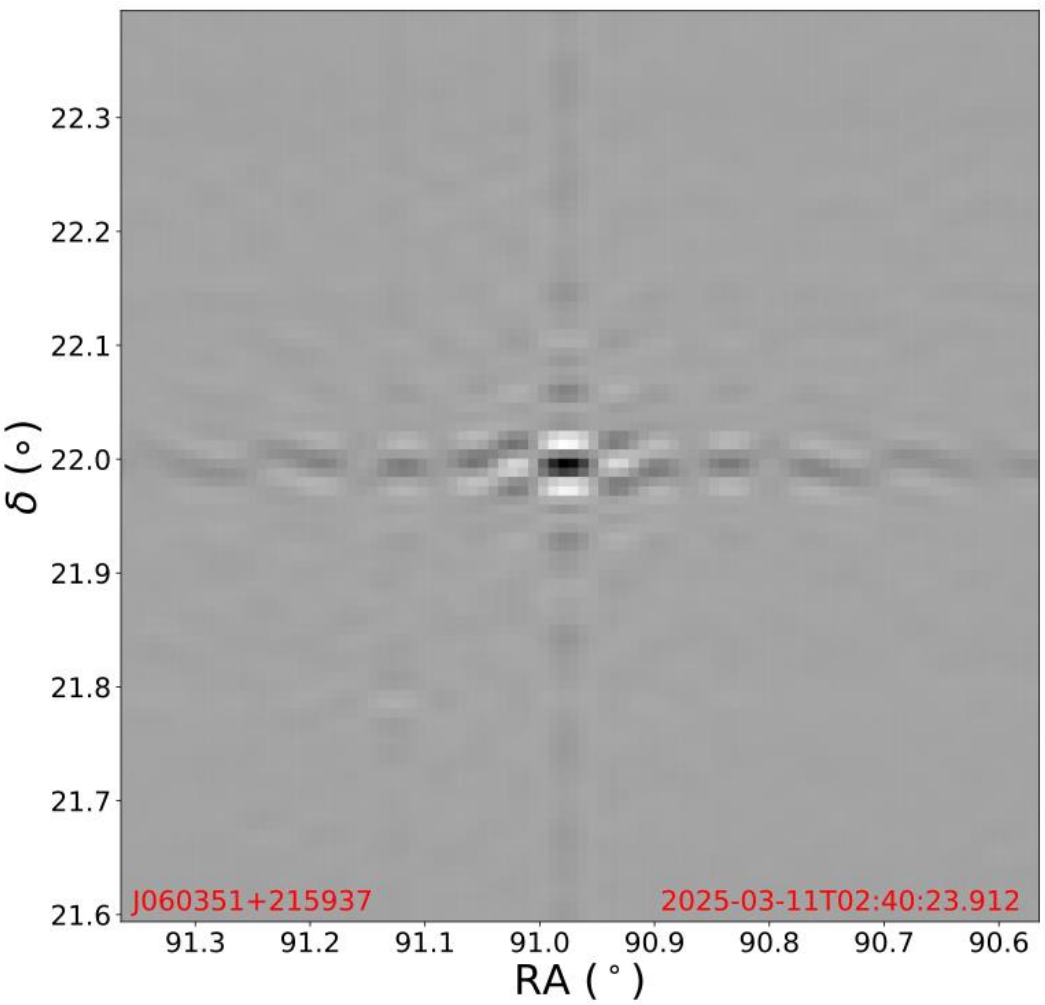}
    \caption{A Time- (3.35\,s) and Frequency- (187.5\,MHz) Averaged Dirty Image of Calibrator J060351+215937 where PSF and central sidelobes are visible. Dark regions are positive and light regions are negative in arbitrary units. J060351+215937 was the brightest calibrator observed during the DSA-110 NSFRB Galactic Plane Survey (DN-GPS) and was used to estimate the flux curve shown in Figure~\ref{fig:calplot}.}
    \label{fig:calplot2}
\end{figure}

To estimate the flux sensitivity we convert the running standard deviation accumulated during the DN-GPS to flux units using 134\,ms single-sample images of the 16 calibrators. The brightest source, J060351+215937 shown in Figure~\ref{fig:calplot2}, is used to compute a conversion factor from arbitrary image pixel units to the NVSS catalog flux (2771.6\,mJy). The same factor is applied to each calibrator observation as shown on the x-axis in Figure~\ref{fig:calplot} (left). Applying this to an observing time-weighted average of the running standard deviation from each observing day, we estimate a sensitivity limit $\sigma_{\rm srch}\approx50$\,mJy, confirming that the calibrators follow an approximately linear relation down to the $3\sigma_{\rm srch}$ flux limit. Note the sensitivity ranged from $30-100\,$mJy depending on the observation date. NVSS and RFC sources are observed daily during real-time observations to continuously update flux estimates; we include real-time NVSS observations in Figure~\ref{fig:calplot} which show that fainter sources' measured flux flattens near $\sigma_{\rm srch}$ as expected.

As indicated in Figure~\ref{fig:sens}, the measured $\sigma_{\rm srch}=50\,$mJy sensitivity exceeds the theoretical radiometer sensitivity estimate with approximately uniform $\mathcal{R}=-2$ image weights $\sigma_{\rm th}=40\,$mJy by roughly $1.3\times$. We attribute this in part to RFI excision. Both far- and near-field RFI have been observed, the former resulting from aircraft passing over the observatory and the latter from cellular networks transmitting near the center of the band. Near field RFI contributes to image-plane noise because it appears at all spatial scales rather than as a point source. The current pipeline adopts two visibility flagging criteria on the 1.5\,MHz channels: (1) the mean visibility in the channel exceeds $5\times$ the running mean, (2) the peak median-subtracted visibility in the channel exceeds $10\times$ the running mean. These have proven effective in excising far-field RFI on a channel basis; however, near-field RFI may be leaking into adjacent channels, biasing the noise high in addition to the bandwidth reduction from channel flagging. Additional analysis is required to fully characterize RFI and devise techniques to recover sensitivity more robustly. Other potential effects that may lower sensitivity include primary beam attenuation, intrinsic flux calibrator variability, and poorly fit beamformer weights on a given day, which we do not investigate in detail here.

\subsection{Results}\label{sec:results}

No candidates were detected by the NSFRB pipeline from the DN-GPS data using a $6\sigma$ threshold. The data were re-searched multiple times with lower thresholds (as low as $2\sigma$) in order to inspect low S/N candidates in more detail, resulting in no marginal candidates. {From the injection test in Section~\ref{sec:detrate} we estimate the 90\% completeness flux limit to be $25\sigma\approx1200\,$mJy.}

As shown in Figure~\ref{fig:burstrate} (bottom), we  partition the survey region into a grid of $3^\circ\times3^\circ$ cells covering the total $\sim774$\,square\,degree survey area to estimate limits on the burst rate $R_{\rm 95}$ from the NSFRB non-detection. Note that we neglect primary beam effects for this calculation. To do this, we first assume that {$R_{\rm 95}<(\hat{f}T_{\rm obs})^{-1}$}, where $T_{\rm obs}=5$\,minutes is the duration of a single pointing dataset, approximately the time for which a target remains in the DSA-110 NSFRB image field. {$\hat{f}$ is the asymptotic detection rate shown in Figure~\ref{fig:finalinjections}, or the fraction of bursts we expect to be detected for high S/N, which we apply as an effective reduction in integration time.} In this limit, we can use Poissonian statistics, adopting the weighted sum observing time from pointings that fall within each region, $T_{\rm tot,eff}$, using noise estimates from the date of observation ($\sim40-60$\,mJy) as inverse weights.

We estimate the upper limit on the total burst rate in each cell as {$R_{\rm 95}=-(\hat{f}T_{\rm tot,eff})^{-1}{\rm ln}(0.05)$} with 95\% confidence. The results are shown in Figure~\ref{fig:burstrate}; we find that $R_{\rm 95}$ ranges from {$3-160$\,hr$^{-1}$}, with a weighted median {$R_{\rm 95}<14.1$\,hr$^{-1}$}. The tightest constraints are at low Galactic latitudes $b$ and at declinations with the most complete coverage. For $|b|<0.25^\circ$, the maximum burst rate upper limit among $3^\circ\times3^\circ$ cells is {$R_{\rm 95}<17$\,hr$^{-1}$} (or {$<2$\,hr$^{-1}$}\,per\,square\,degree).

\section{Discussion}\label{sec:discussion}

While no candidates are detected, this is the first targeted search for LPRTs at 1.4 GHz, and the burst rate limit can be used to constrain models for LPRT emission. In this section, we contrast the detectability of two predominant models: the isolated neutron star model from \citet{cooper2024beyond} and the White Dwarf-M Dwarf model from \citet{qu2025magnetic}. 

\subsection{Relevant Noise Estimates for Long and Short Periods}

For this discussion, we partition the analysis at pulse period {$P=\hat{f}T_{\rm obs}=\hat{f}\times(5$\,minutes$)=2\,$minutes; for $P<\hat{f}T_{\rm obs}$}, we assume that the LPRT is always ``on", thus, if the flux exceeds the {90\% completeness threshold $25\sigma$ (see Section~\ref{sec:detrate}}) in any $5-$minute observation, it will be detected. Therefore, for each $3^\circ\times3^\circ$ grid cell, we take the minimum noise estimate among each contributing observing day as the relevant detection limit. The median among all grid cells is then {$\sigma_{\rm srch,P<2\,min}\approx 40\,$mJy}. 

For {$P>\hat{f}T_{\rm obs}$}, although the source is always ``on", the long period allows us to treat it as a Poissonian process. {Consider a source with {$P>\hat{f}T_{\rm obs}$}; since the source remains within the DSA-110 field-of-view ($\sim1.5^\circ\,\cos(\delta)^{-1}$) for at most $\sim5$ minutes (at the upper declination limit $\delta\approx65^\circ$), only one pulse at most can be detected per day. The phase offset of this first pulse is effectively random since we have no prior information about the phase; subsequent bursts in contrast must be phase-connected to the first. Therefore, as we only perform a single-pulse search, the arrival of the first pulse is similar to a Poissonian random process with burst rate $\mathcal{R}\approx1/P$.} Just as the effective rate limit adopts a weighted sum observing time, $T_{\rm tot,eff}$, for each grid cell, we use the median noise estimate among each contributing observing day as the relevant detection limit. The median among all grid cells is then {$\sigma_{\rm srch,P>2\,min}\approx 50\,$mJy. In the next section we use $\sigma_{\rm srch} = \sigma_{\rm srch,P<2\,min}$ and $\sigma_{\rm srch} = \sigma_{\rm srch,P>2\,min}$ for $P<2$\,minutes and $P>2$
\,minutes} unless otherwise stated. The distinction is indicated in Figure~\ref{fig:constraint} by a labeled vertical grey line at {$P=2\,$minutes}.

{In the following analysis, we assume the LPRT sources are `always active' for simplicity; some sources have appeared active for only months at a time \citep[e.g.][]{hurley2022radio} while others have been active for decades \citep[e.g.][]{hurley2023long}. Therefore, we leave a more detailed treatment of activity windows to future work.}

\subsection{Constraints on the White Dwarf-M Dwarf Binary and Neutron Star Models}

\begin{figure*}
    \centering
    \includegraphics[width=\textwidth]{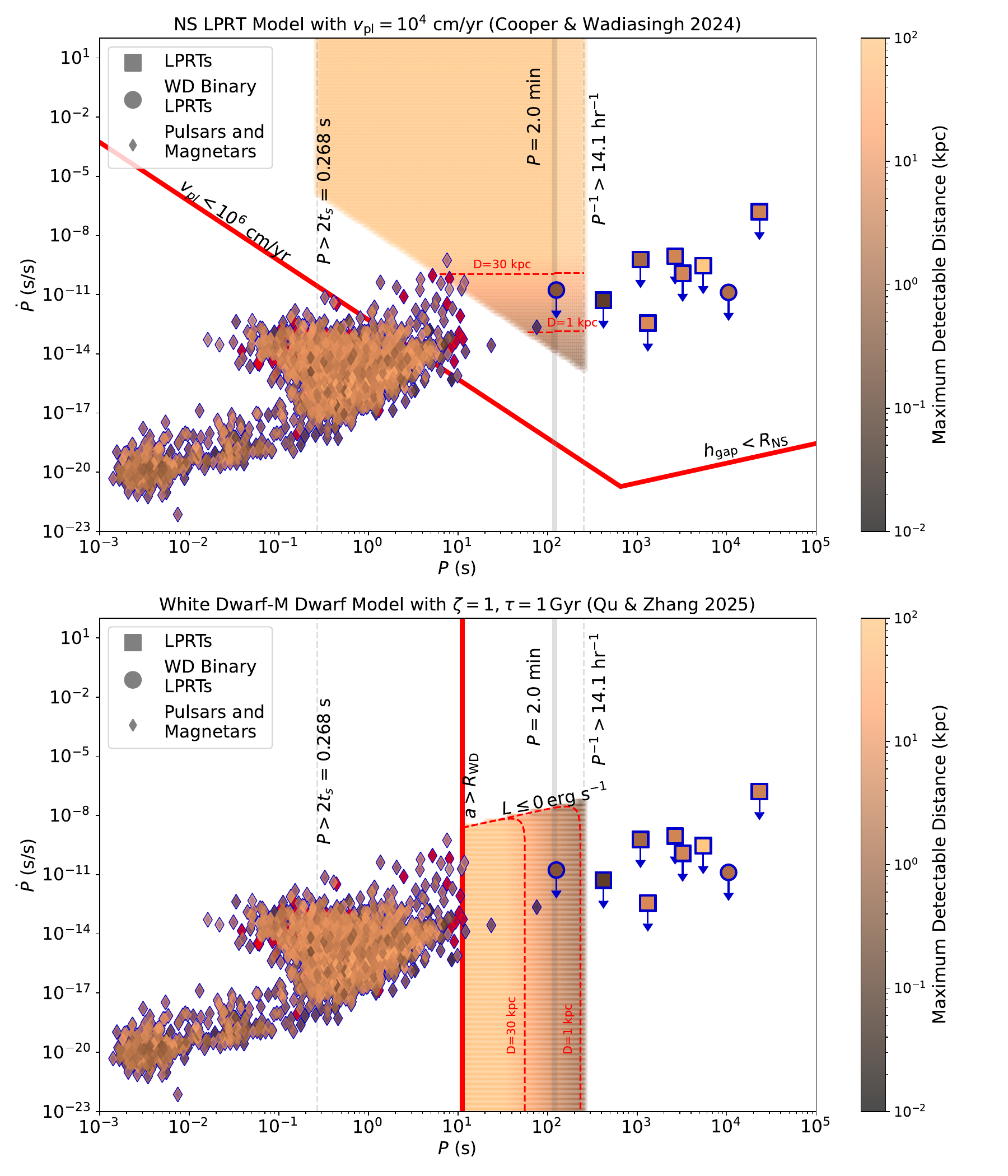}
    \caption{$P-\dot{P}$ Constraints from the DN-GPS Survey Using a Neutron Star Model \citep[top;][]{cooper2024beyond} and a White Dwarf Model \citep[bottom;][]{qu2025magnetic}. In each plot the thick red lines bound the allowed emission ranges for each model. Vertical grey lines indicate hard limits from the DN-GPS search due to the sampling time and Poissonian burst rate. Pulsars (diamonds), WD binary LPRTs (squares) and all other LPRTs (circles) are colored with their distance estimates, while those without distances are in red. The colorbar also highlights the maximum detectable distance ($d_{\rm max}(P,\dot{P})$) for each model (see Section~\ref{sec:ngps}) given the DN-GPS {$25\sigma_{\rm srch,P<5\,min}\approx1030$\,mJy and $25\sigma_{\rm srch,P>5\,min}\approx1200$\,mJy} detection limits for {$P<2\,$min and $P>2$\,min}, respectively. Red dashed lines are contours of constant $d_{\rm max}(P,\dot{P})$).}
    \label{fig:constraint}
\end{figure*}

In the \citet{cooper2024beyond} neutron star model, plastic motion of the crust of a post-deathline magnetar induces twist in the magnetic field, storing magnetic energy. This energy is released over $\sim$days, driving pair-plasma creation of charged particles. The resulting charges accelerate in a vacuum gap potential and coherently radiate via curvature radiation or inverse-Compton scattering. The maximum dissipation luminosity, which notably excludes the efficiency of pair-creation \citep[e.g.][]{arendt2002pair}, is below:

\begin{equation}
    L_{\rm diss,NS} \approx \frac{v_{\rm p}B_S^2A_{\rm fp}R_{NS}}{cP\sin^2(\theta_{\rm fp})}
\end{equation}

\noindent where $v_{\rm p}\lesssim10^6\,$cm\,yr$^{-1}$ is the velocity of plastic flow, $B_S\approx10^{15}$\,G is the surface magnetic field strength, $A_{\rm fp}\approx10\,$km is the size of the crust `footprint" undergoing plastic motion, and $\theta_{\rm fp}$ is the co-latitude of the footprint. By asserting that {$L_{\rm diss,NS}>25\sigma_{\rm srch}d^2{\rm BW}$} for an LPRT at distance $d$ within this model, and using the characteristic $B_S\gtrsim (3.2\times10^{19}\,G)\sqrt{P\dot{P}/2}$, we can solve for the maximum distance $d_{\rm max,NS}(P,\dot{P})$ to which the DN-GPS could detect the LPRT. Figure~\ref{fig:constraint} (top) shows this result with $v_{\rm pl}=10^4$\,cm\,yr$^{-1}$; we see that for {$\dot{P}>10^{-13}$} and {$P<R^{-1}_{\rm 95}=4.3$\,min} the search is complete to only 1\,kpc. To the edge of the galaxy, approximately $30\,$kpc, the search is only sensitive to {$\dot{P}>10^{-10}$}. Based on the $\dot P$ range of detected LPRTs with {$P<R^{-1}_{\rm 95}=4.3$\,min}, this indicates that the search is not sensitive to magnetar-LPRTs beyond $1\,$kpc. In addition, \citet{cooper2024beyond} indicate that the plastic flow velocity can reasonably range from $1<v_{\rm pl}<10^6$\,cm\,yr$^{-1}$, which is proportional to $L_{\rm diss,NS}$ and can significantly alter the horizon distance. Therefore, the search is not complete for magnetar-LPRT objects; we conclude that a population of ultra-long period magnetars may still be present within the Galactic Plane despite our non-detection.


In the \citet{qu2025magnetic} White Dwarf-M Dwarf binary model, the companion star orbits within the White Dwarf's magnetosphere, creating a voltage drop between the two stars via the unipolar inductor effect (Faraday disk). As charge flows between the stars due to the potential drop they create an electrical current, and the magnetosphere dissipates power via Ohmic heating, driving pair creation. Pair charges then radiate via the relativistic electron cyclotron maser mechanism. We adjust the derivation from \citet{qu2025magnetic} to allow non-zero spin-down $\dot P$; a brief derivation is provided in Appendix~\ref{app:deriv}, obtaining the following dissipation luminosity:

\begin{equation}\label{eq:LdissWD}
    L_{\rm diss,WD}  = \frac{4\pi}{c}\biggl(\frac{ B_{\rm WD}R_{\rm WD}^3R_c\zeta(4\pi^2)^{2/3}}{(G(M_{\rm WD}+M_c))^{2/3}}\biggr)^2(\frac{1}{P^{14/3}} - \frac{14}{3}\frac{\dot{P}\tau}{P^{17/3}})
\end{equation}

\noindent where $B_{\rm WD}\approx10^{6}$\,G is the surface magnetic field strength of the White Dwarf star, $M_{WD}$ and $M_{c}$ are the stars' masses, $R_{WD}$ and $R_{c}$ are the stars' radii, and $\tau\lesssim1$\,Gyr is the timespan over which the White Dwarf radiates. Here we normalize $\zeta\equiv|P^{-1}-P_{\rm rot}^{-1}|P= 1$ where $P$ is the orbital period and $P_{\rm rot}$ is the White Dwarf's rotation period; we assume $\tau\approx\tau_{\rm age}=1\,$Gyr, the typical White Dwarf age, as an upper limit, which will yield a more conservative range of $\dot{P}$ values (see Appendix~\ref{app:deriv} for justification of these assumptions). We apply the same detection condition, {$L_{\rm diss,WD}>25\sigma_{\rm srch}d^2{\rm BW}$}, and the resulting $d_{\rm max,WD}(P,\dot P)$ is shown in Figure~\ref{fig:constraint} (bottom). In this case, the DN-GPS is sensitive to WD-binary-LPRT objects within $d\lesssim 30\,$kpc for {$P\lesssim50\,$s}. By this model, our non-detection determines that across {720\,square\,degrees} ($\sim95\%$) of the survey region (excluding only {6} grid-points on the outskirts with {$R_{\rm 95}^{-1}<50$\,s}) there are no WD-binary-LPRTs with {$S_{\rm 1.4\,GHz}>25\sigma_{\rm srch,P<2\,min}\approx1030$\,mJy and $10\lesssim P\lesssim50$\,s}. While only a fraction of the allowed $P-\dot{P}$ parameter space is constrained, our results motivate further discussion of both the White Dwarf-M Dwarf binary and magnetar LPRT models.

\section{Conclusion}\label{sec:conclusion}

We have developed and commissioned the DSA-110 NSFRB instrument, a GPU-accelerated image-plane single-pulse search sensitive to bright LPRTs with pulse widths between $134\,$ms$-160.8\,$s. Initial continuum source observations have been largely successful and the pulsar B0329+54 is routinely detected. The inaugural DN-GPS survey recovered no candidates, placing limited constraints on the White Dwarf-M Dwarf binary model. The real-time NSFRB search is online in drift-scan mode, adopting hourly source injections, daily calibrator and pulsar observations, and weekly construction of classifier re-training sets from false-positive detections for system health evaluation. Future work will explore triggered periodicity searching, further expansion of the pulse width range, and more DN-GPS-like surveys of the Galactic Plane, star-forming regions, and globular clusters. 

\section{Data Availability}

The NSFRB code is publicly available at \url{https://github.com/dsa110/dsa110-nsfrb}. 

\begin{acknowledgments}

The authors would like to thank Adam Dong, Maura McLaughlin, Duncan Lorimer, Scott Ransom, Bing Zhang, Tony Rodriguez, Assaf Horesh, Jayanta Roy, Bhaswati Bhattacharyya, Emmanuel Fonseca, Graham Doskotch{, Natasha Hurley-Walker, and Kshitij Bane} for creative discussions exploring the unique nature of LPRTs and novel ways to search for them. {The authors thank the anonymous referee for offering constructive comments that improved the quality of the final manuscript.} This work includes data from the Radio Fundamental Catalogue (RFC; \url{www.doi.org/10.25966/dhrk-zh08}) This material is based upon work supported by the National Science Foundation Graduate Research Fellowship under Grant No. 2139433. The authors thank staff members of the Owens Valley Radio Observatory and the Caltech radio group, whose tireless efforts were instrumental to the success of the DSA-110. The DSA-110 is supported by the National Science Foundation Mid-Scale Innovations Program in Astronomical Sciences (MSIP) under grant AST-1836018.

\end{acknowledgments}





\appendix


\section{Derivation of Time-dependent Dissipation Luminosity for the White Dwarf Model}\label{app:deriv}

In this appendix, we derive equation~\ref{eq:LdissWD} incorporating time dependence. Starting from Equation\,9 in \citet{qu2025magnetic}, the total voltage when the White Dwarf and M-dwarf are at a separation $a(t)$ is given by:

\begin{equation}
    \Phi(t) = \frac{2\mu R_c}{ca(t)^2}\zeta \Omega(t)
\end{equation}

\noindent where we introduce time-dependence in the orbital period $P=2\pi/\Omega$ and orbital separation $a$, and define $\zeta\equiv|\Omega-\Omega_{\rm rot}|/\Omega \sim $\,constant where $\Omega_{\rm rot}$ is the rotational period of the White Dwarf. Recall the magnetic dipole moment $\mu \approx B_{\rm WD}R_{\rm WD}^3$, and Kepler's law $a(t)^3=G(M_{\rm WD}+M_{\rm c})P(t)^2/(4\pi)$. Plugging these in for $\mu$ and $a(t)$, respectively:

\begin{equation}
    \Phi(t) = \frac{2B_{\rm WD}R_{\rm WD}^3R_c\zeta(4\pi^2)^{2/3}}{c(G(M_{\rm WD}+M_c)P^2(t))^{2/3}}\frac{2\pi}{ P(t)}
\end{equation}

\begin{equation}
    \Phi(t) = \frac{2B_{\rm WD}R_{\rm WD}^3R_c\zeta(4\pi^2)^{2/3}}{c(G(M_{\rm WD}+M_c))^{2/3}}\frac{2\pi}{ P^{7/3}(t)} = KP(t)^{-7/3}, \quad K\equiv \frac{4\pi B_{\rm WD}R_{\rm WD}^3R_c\zeta(4\pi^2)^{2/3}}{c(G(M_{\rm WD}+M_c))^{2/3}}
\end{equation}

\noindent Let the resistance of the magnetosphere be $R_{\rm mag}=4\pi/c$. The ohmic dissipation power is then written as:

\begin{equation}
    \dot{E}_{\rm diss}(t) = \frac{c}{4\pi}K^2P(t)^{-14/3}
\end{equation}

\noindent If we assume the orbital period evolves slowly as $P(t) = P_0 + \dot{P}t$ to first order, we can write the dissipation power as:

\begin{equation}
    \dot{E}_{\rm diss}(t) = \frac{c}{4\pi}K^2(P_0 + \dot{P}t)^{-14/3}
\end{equation}

\begin{equation}
    \dot{E}_{\rm diss}(t) = \frac{c}{4\pi}K^2P_0^{-14/3}(1 + \frac{\dot{P}}{P_0}t)^{-14/3}
\end{equation}



\noindent If $\dot{P}t\ll P_0$, we can expand this to first order:

\begin{equation}
    \dot{E}_{\rm diss}(t) = \frac{c}{4\pi}K^2P_0^{-14/3}(1 - \frac{14}{3}\frac{\dot{P}}{P_0}t)
\end{equation}

\noindent Then, substituting in for $K$ and evaluating at the timespan over which the White Dwarf radiates $t=\tau$, we obtain equation~\ref{eq:LdissWD}:

\begin{equation}
    L_{\rm diss,WD} \equiv \dot{E}_{\rm diss}(\tau) = \frac{4\pi}{c}\biggl(\frac{B_{\rm WD}R_{\rm WD}^3R_c\zeta(4\pi^2)^{2/3}}{(G(M_{\rm WD}+M_c))^{2/3}}\biggr)^2(\frac{1}{P_0^{14/3}} - \frac{14}{3}\frac{\dot{P}\tau}{P_0^{17/3}})
\end{equation}

\noindent Note that in general, we expect $\tau\ll\tau_{\rm age}\sim1$\,Gyr; for example, GPM\,J1839-10 was detected in archival data over the course of 36\,years. Using its upper limit $\dot{P}<3.6\times10^{-13}$ and measured $P=31482.4\,$s, we find $\dot{P}\tau/P\sim10^{-8}$ for $\tau =36\,$years \citep{horvath2025unified,men2025highly}. Even for $\tau=1\,$Gyr, we have $\dot{P}\tau/P\sim10^{-1}$. Therefore we conclude the solution is valid in general for the LPRT case. Note also that $\zeta$ can vary over a wide range ($\zeta\sim22$ for GPM\,J1839-10), but we normalize it to unity for this discussion as in \citet{qu2025magnetic}.

\section{RFI Examples}\label{app:rfi}

\begin{figure*}
    \centering
    \includegraphics[width=\textwidth]{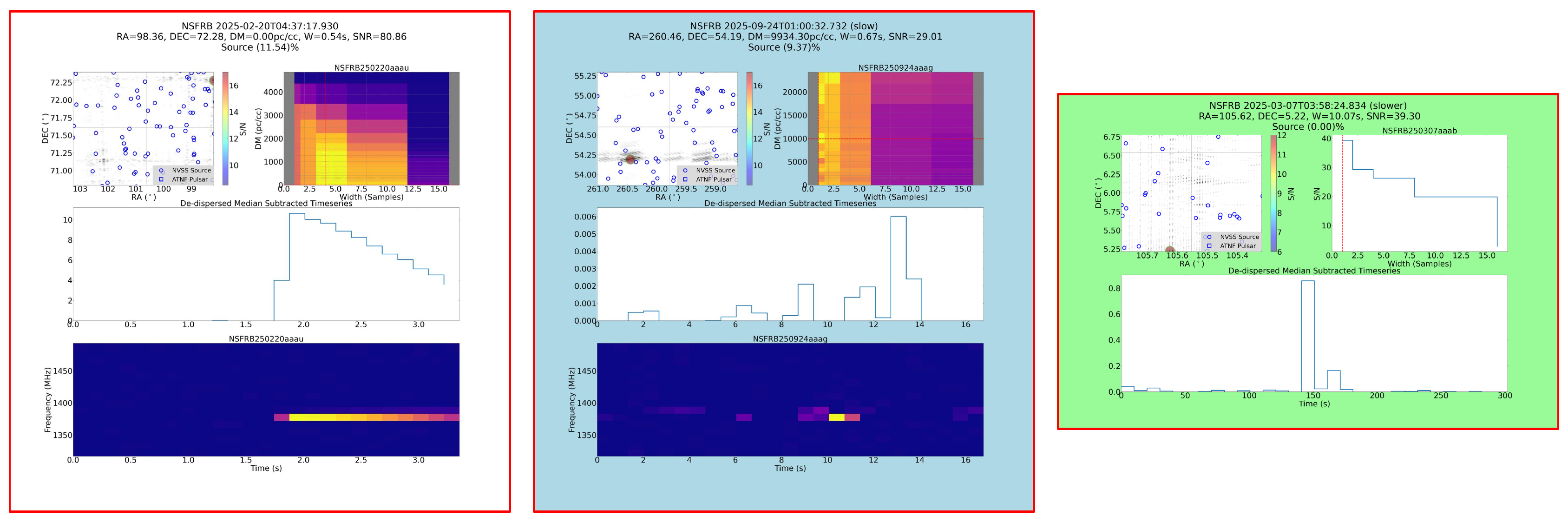}
    \caption{{Candidate Plots for RFI Detected in the 134\,ms (left) 670\,ms (middle) and 10.07\,s (right) Searches. Subplots for the left and middle plots are the same as described in Figure~\ref{fig:b0329}, while the right plot omits the dynamic spectrum since de-dispersion is omitted for the 10.07\,s search.}}
    \label{fig:rfiplots}
\end{figure*}

{Figure~\ref{fig:rfiplots} shows example candidate plots for RFI on each search timescale, which was later added to the training set for the CNN classifier.}

\bibliography{PASPsample701}{}
\bibliographystyle{aasjournalv7}



\end{document}